\documentclass[prl,superscriptaddress,a4paper,twocolumn]{revtex4-1}
\usepackage{geometry}
\usepackage[english]{babel}
\usepackage[latin1]{inputenc}
\usepackage{hyperref}
\usepackage{amsfonts}
\usepackage{amsmath, amssymb, amsfonts, mathrsfs}
\usepackage{graphicx}
\usepackage{xspace}

\usepackage{multirow}
\usepackage[bottom]{footmisc}

\newcommand*{\melvin}{{\small M}{\scriptsize ELVIN}\xspace}

\begin{document}

\title{Gouy Phase Radial Mode Sorter for Light: Concepts and Experiments}

\author{Xuemei Gu}
\email{xmgu@smail.nju.edu.cn}
\affiliation{Institute for Quantum Optics and Quantum Information (IQOQI), Austrian Academy of Sciences, Boltzmanngasse 3, 1090 Vienna, Austria.}
\affiliation{State Key Laboratory for Novel Software Technology, Nanjing University, 163 Xianlin Avenue, Qixia District, 210023, Nanjing City, China.}

\author{Mario Krenn}
\email{mario.krenn@univie.ac.at}
\affiliation{Institute for Quantum Optics and Quantum Information (IQOQI), Austrian Academy of Sciences, Boltzmanngasse 3, 1090 Vienna, Austria.}
\affiliation{Vienna Center for Quantum Science \& Technology (VCQ), Faculty of Physics, University of Vienna, Boltzmanngasse 5, 1090 Vienna, Austria.}

\author{Manuel Erhard}
\affiliation{Institute for Quantum Optics and Quantum Information (IQOQI), Austrian Academy of Sciences, Boltzmanngasse 3, 1090 Vienna, Austria.}
\affiliation{Vienna Center for Quantum Science \& Technology (VCQ), Faculty of Physics, University of Vienna, Boltzmanngasse 5, 1090 Vienna, Austria.}

\author{Anton Zeilinger}
\email{anton.zeilinger@univie.ac.at}
\affiliation{Institute for Quantum Optics and Quantum Information (IQOQI), Austrian Academy of Sciences, Boltzmanngasse 3, 1090 Vienna, Austria.}
\affiliation{Vienna Center for Quantum Science \& Technology (VCQ), Faculty of Physics, University of Vienna, Boltzmanngasse 5, 1090 Vienna, Austria.}

\begin{abstract}
We present an in principle lossless sorter for radial modes of light, using accumulated Gouy phases. The experimental setups have been found by a computer algorithm, and can be intuitively understood in a geometric way. Together with the ability to sort angular-momentum modes, we now have access to the complete 2-dimensional transverse plane of light. The device can readily be used in multiplexing classical information. On a quantum level, it is an analog of the Stern-Gerlach experiment -- significant for the discussion of fundamental concepts in quantum physics. As such, it can be applied in high-dimensional and multi-photonic quantum experiments.

\end{abstract}
\date{\today}
\maketitle
\section{Introduction}
The spatial modes of light give access to an in principle unbounded state space. This allows to encode information in higher-dimensional alphabets beyond one bit per photon. A particularly well-studied mode family are the Laguerre-Gaussian (LG) modes \cite{padgett2017orbital}. One of the special features of LG modes is that they carry well defined $\ell\hbar$ quanta of orbital-angular momentum (OAM) or $\ell$ is one of the two numbers that define the LG modes \cite{allen1992orbital}. Over the last 25 years, a large toolbox has been developed to generate \cite{marrucci2006optical, campbell2012generation, fickler2016quantum} and manipulate \cite{leach2002measuring, berkhout2010efficient, lavery2012refractive, mirhosseini2013efficient, zhang2016engineering, babazadeh2017high} OAM modes. This has led to a manifold of applications ranging from classical high-speed \cite{wang2012terabit, huang2014100} and long-distance \cite{krenn2016twisted, lavery2017free} communication to high-dimensional quantum entanglement \cite{mair2001entanglement, dada2011experimental, romero2012increasing, vaziri2002experimental} and quantum cryptography \cite{groblacher2006experimental, mirhosseini2015high, sit2017high}.

The second much less investigated degree of freedom of the LG modes is the radial quantum number $p$ \cite{karimi2012radial, karimi2014radial, plick2015physical}. It spans a second completely independent and in principle unbounded state space, with the same capability to encode a vast amount of information. Its quantum character has been demonstrated by two-photon interference  \cite{karimi2014exploring} and quantum correlations and entanglement between radial modes have been demonstrated \cite{salakhutdinov2012full, krenn2014generation}. The only real manipulation of radial modes known so far has been demonstrated using the potential of controlled random material, in order to sort different radial modes \cite{fickler2017custom}. Unfortunately the manipulation works in a lossy way which made its application in classical and quantum experiments challenging so far. In order to exploit the full potential of the radial modes, the available toolbox needs to be expanded.

Here we show how higher-order spatial modes of light -- in particular their radial modes -- can be sorted with theoretically 100\% visibility. For that, we use an interferometer with a mode-dependent phase difference. This concept has been used for other degrees-of-freedom (such as a mode sorter for OAM modes in \cite{leach2002measuring}, or in a theoretical proposal for a more general interferometric sorting scheme \cite{ionicioiu2016sorting}). The challenging question then is: \textit{How can one experimentally achieve mode dependent phase-shifters for radial modes?} We answer this question with the help of a computer algorithm \cite{krenn2016automated}: One of the two arms in the interferometer contains a lens configuration, which leads to a difference in the accumulated Gouy phase $\Delta\varphi_{g}$. We show experimentally a phase difference of $\Delta \varphi_{g}=\frac{\pi}{2}$ and $\Delta \varphi_{g}=\frac{\pi}{4}$, and apply it to sort different spatial modes. We also theoretically demonstrate phase differences of $\Delta \varphi_{g}=\frac{\pi}{n}$ ($2 \leq n \leq 8$).

Our technique can readily be applied to classical experiments (such as (de)multiplexing in classical communication) or quantum experiments (such as two-photon interference for high-dimensional entanglement). Our intuitive geometric interpretation of the technique can be used to unify and generalize several similar approaches.

\begin{figure*}
\includegraphics[width=\textwidth]{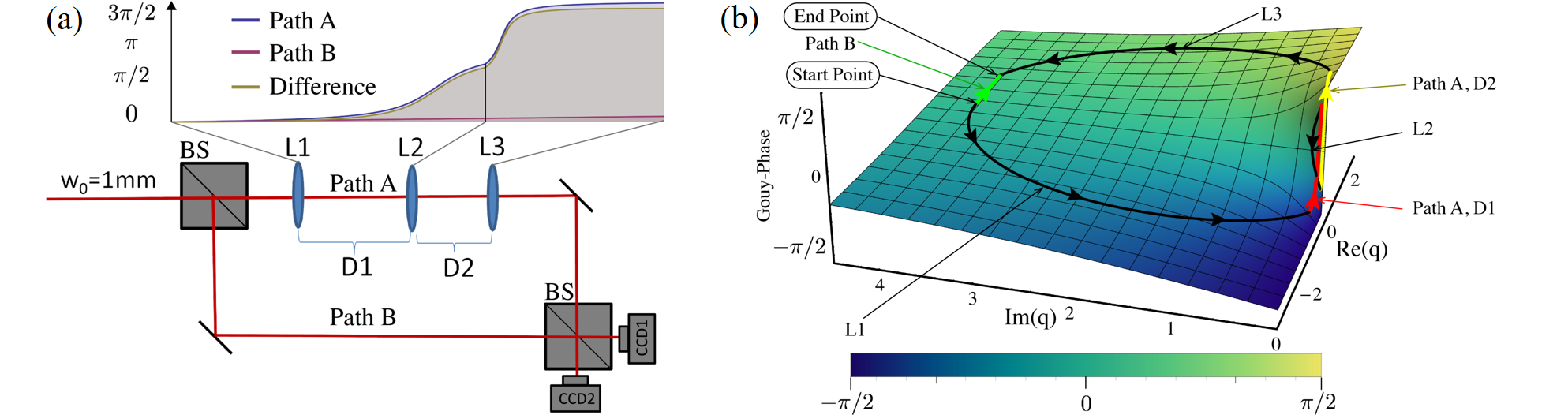}
\caption{Accumulated Gouy Phase Interferometer. \textbf{A}:
Schematic of the theory setup with accumulated Gouy phase. A Mach-Zehnder interferometer with three lenses (L1, L2, L3 : 500, 40, 300mm; D1, D2 : 560.8, 344.3mm for 1mm collimated incoming beam) in one arm. The phase difference is introduced by the beam propagated through the three lenses. We obtain the accumulated Gouy phase difference $\Delta\varphi_{g}=\frac{3\pi}{2}=-\frac{\pi}{2}$. The device can sort light beams depending on the $p$ and $\ell$ values, which will be detected through the two output CCD1 or CCD2. \textbf{B}: Complex $q$-parameter space: The complex beam parameter $q$ determines the spatial shape of a Gaussian beam -- its beam waist and radius of curvature -- unambiguously. The Gouy phase of a beam can be determined as a function of $q$, which is the plotted surface. The accumulated Gouy phase of a beam while propagating freely is given by height difference on the surface. Here, the green line corresponds to the free propagation in path B, while the red and yellow line corresponds to the propagation after L1 and after L2, respectively. The black lines indicate the action of the lenses in the $q$-space: Both $z$ and $z_R$ change discretely, which gives a jump in the $q$-space. The shape of the black curves is calculated from a continuous transition of the lense from f=$\infty$ to f=500mm, f=40mm and f=300mm, respectively. In order to interfere, the two beams in path A and B must have the same spatial properties, i.e. must have the same $q$ parameter. Thus, the endpoint of the green line must be at the endpoint of the black line for lens L3.}
\label{fig:gouyphase}
\end{figure*}

\textbf{Spatial modes and Gouy Phase} -- The paraxial wave equation in cylindric coordinates leads to Laguerre-Gaussian mode solutions, which are denoted as
\begin{align*}
LG&_{p,\ell}(r,\phi,z)=\\\nonumber
&=\sqrt{\frac{2p!}{\pi(p+\left | \ell \right |)!}}\frac{1}{w_{z}}\left (\frac{\sqrt{2}r}{w_{z}}\right)^{\left| \ell \right|}L_{p}^{\left | \ell \right |} \left( \frac{2 r^{2}}{w_{z}^{2}}\right)\\\nonumber
&\times \exp\left(-\frac{r^{2}}{w_{z}^{2}} + i\left( \frac{kr^{2}}{2R_{z}}-k z + \ell\phi-(2p+\left | \ell \right |+1)\varphi_{g} \right) \right)
\label{lightbeam}
\end{align*}
with the mode number $\ell$ denoting the orbital angular momentum (in units of $\hbar$) and $p$ is a radial mode number. $L_{p}^{\left| \ell \right|}$ are the Laguerre polynomials, $w_{z}=w_{0}\sqrt{1+(\frac{z}{z_{R}})^{2}}$ is the beam waist with $w_0$ being the beam waist at the focus. $z_{R}=\frac{\pi w_0^2}{\lambda}$ is the Rayleigh range, $R_{z}=z\left(1+\left(\frac{z}{z_{R}}\right)^{2}\right)$ is the radius of curvature, $\lambda$ is the wavelength and $k=\frac{2\pi}{\lambda}$ is the wave number. $\varphi_{g}=\arctan(\frac{z}{z_{R}})$ denotes the Gouy phase and is multiplied with the mode order $m=(2p+|\ell|+1)$.

This phase is accumulated by Gaussian beams when they propagate through the focus, and has first been observed by Louis Georges Gouy in 1890. Several physical interpretations have been given, such as the geometric effect of the gaussian beam \cite{boyd1980intuitive}, as a geometric \cite{simon1993bargmann} or topological phase \cite{subbarao1995topological}, or a phenomena arising due to an uncertainty relation \cite{feng2001physical}. The Gouy phase has been used with higher-order Gaussian modes, for example to convert between higher-order modes \cite{beijersbergen1993astigmatic} or to interferometrically sort Hermite-Gauss modes \cite{linares2017spatial, linares2017interferometric}.

\begin{figure}[hb]
\includegraphics[width=0.5\textwidth]{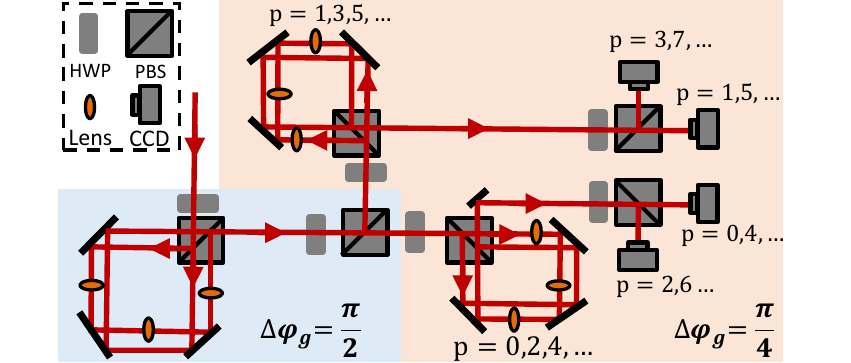}
\caption{To sort higher mode-numbers, one could cascade the interferometers (a folded Sagnac-interferometer for long-term stability). Combining three such interferometers allows for sorting modes from $p=0$ to $p=3$: The first interferometer (with blue background) sorts even and odd modes, while the subsequent two interferometers sort $p=0$/$p=2$ and $p=1$/$p=3$, respectively. In our experiments, we show the results for both types of interferometers.}
\label{fig:cascade}
\end{figure}

\begin{figure*}
\includegraphics[width=\textwidth]{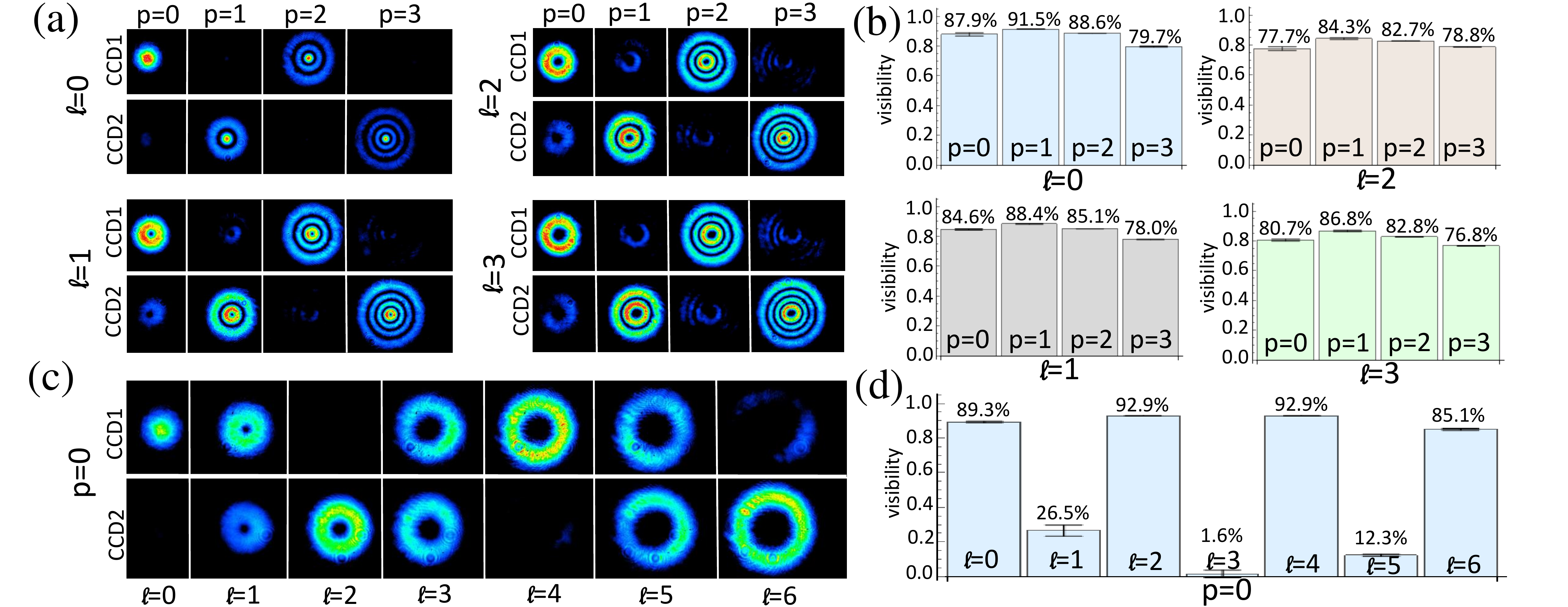}
\caption{Experimental results for $p$ and $\ell$ modes in a $\Delta\varphi_{g}=\frac{\pi}{2}$ interferometer. \textbf{A}: We show the sorting of four different $p$-modes from $p$=0 to $p$=3, with different OAM values of $\ell$=0-3. The intensity images taken from CCD cameras clearly show high quality sorting of different mode orders. To quantify the quality, we calculate the visibility $vis=\left|\frac{P_1 - P_2}{P_1 + P_2}\right|$ (where $P_i$ stands for the power in the output arm $i$). All visibilities are beyond 75\%. The errors stand for statistical uncertainties calculated from 10 independent measurement data points. \textbf{C-D}: We measure OAM modes $\ell$=0 to $\ell$=6. Every even mode number is sorted, while every odd mode number is splitted and propagates to both detectors, which gives a vanishing visibility.}
\label{fig:halfPisort}
\end{figure*}

We use an interferometer, where one of the two arms contains a lens configuration and thus accumulate a different of Gouy phase \cite{erden1997accumulated, arai2013accumulated}, as shown in Fig. \ref{fig:gouyphase}A. When they recombine, the two beams have the same spatial dimensions (beam waist and radius of curvature) but a mode-dependent phase-difference $m\cdot\Delta \varphi_{g}$. If the phases are fractions of $\pi$, the interferometer can be used to sort higher-order Gaussian modes. The action of the interferometer can be intuitively understood in a geometric way, shown in Fig. \ref{fig:gouyphase}B. For that, we take advantage of the \textit{complex beam parameter} $q = z + i z_{R}$, which completely determines the spatial properties of (higher order) Gaussian beams after propagation through a lens system. Thus, also the Gouy phase can be written in terms of $q$ as $\varphi_{g}(q)=\arctan(\frac{\textnormal{Re(q)}}{\textnormal{Im(q)}})$, where $Re()$ and $Im()$ stand for the Real and Imaginary part of the complex $q$ parameter. One can plot the Gouy phase in the complex $q$-space, where the two coordinates are the real and imaginary part of the $q$ parameter. In this space, one can plot the propagation of the two beams, and directly observe the accumulation of the Gouy phase, as shown in Fig. \ref{fig:gouyphase}B. The free-space propagation of a Gaussian beam continuously changes Re($q$), while a lens performs a discrete jump in the complex $q$-space (as it discretely changes $z$ and $z_R$ at the same time). The Gouy phase difference is then the difference between the accumulated phase of the two beams.

By combining several of these interferometers, one can in principle access very high-dimensional mode spaces (see Fig. \ref{fig:cascade}). Cascaded interferometers can be built extremely stable with modern production techniques (such as shown in \cite{wang201818}, where 30 interferometers have been stable for more than 3 days with a visibility of more than 99\%).

\textbf{Experimental implementation} -- We implement the experimental setup shown in Fig. \ref{fig:gouyphase}A in order to sort higher-order Laguerre-Gauss modes, in particular their radial modes. For that, we require that the Gouy phase is a fraction of $\pi$, as well as that the resulting Gaussian beams have the same complex beam parameter $q$ in order to interfere perfectly. With arbitrary, custom-tailored lens configurations, we could easily find such configurations (as it is indicated in Fig. \ref{fig:gouyphase}B). However, we restrict ourselves to standard lenses which are easily commercially available (see Supplementary for a list of lenses). We use the computer algorithm \melvin to search for suitable experimental configurations \cite{krenn2016automated}. We found setups using three lenses in one of the arms, for $\Delta\varphi_{g}=\frac{\pi}{n}$ with $2 \leq n \leq 8$ in the Supplementary, and it is straight forward to find other phase configurations.

Experimentally we realized $\Delta\varphi_{g}=\frac{\pi}{2}$ and $\Delta\varphi_{g}=\frac{\pi}{4}$, where both configurations use the same lenses but different distances $D1$ and $D2$ between them.

\begin{figure}[hb]
\includegraphics[width=0.5\textwidth]{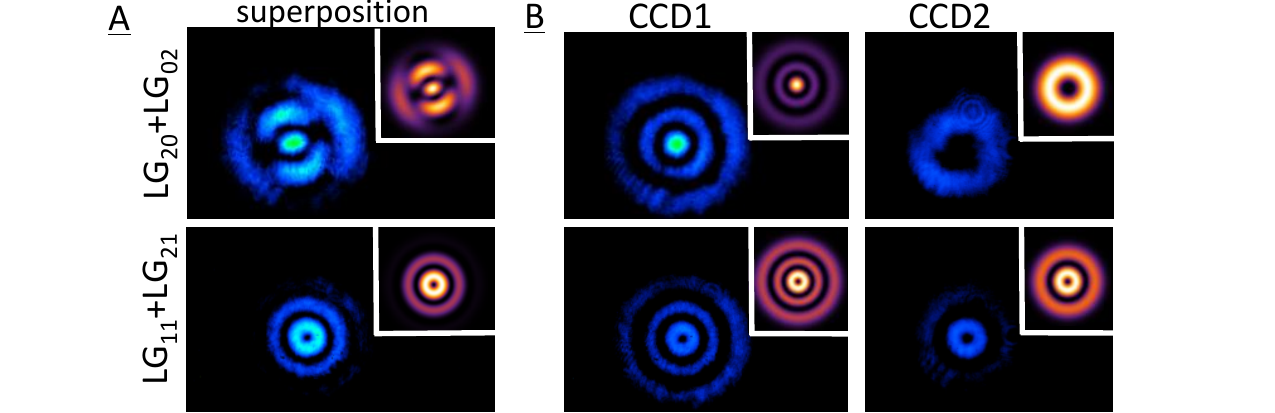}
\caption{Mode superpositions $\psi=\frac{1}{\sqrt{2}}\left(\text{LG}_{2,0}+\text{LG}_{0,2}\right)$ and $\psi=\frac{1}{\sqrt{2}}\left(\text{LG}_{1,1}+\text{LG}_{2,1}\right)$ -- \textbf{A}: The images show the intensity of the superposition, the inset shows the theoretical intensity distribution. \textbf{B}: Measured intensity distributions. The following two images show the output of the CCDs in the two output arms. As the modes have different mode numbers, they are sorted to different outputs of the interferometer.}
\label{fig:Superposition}
\end{figure}

\begin{figure*}[ht]
\includegraphics[width=\textwidth]{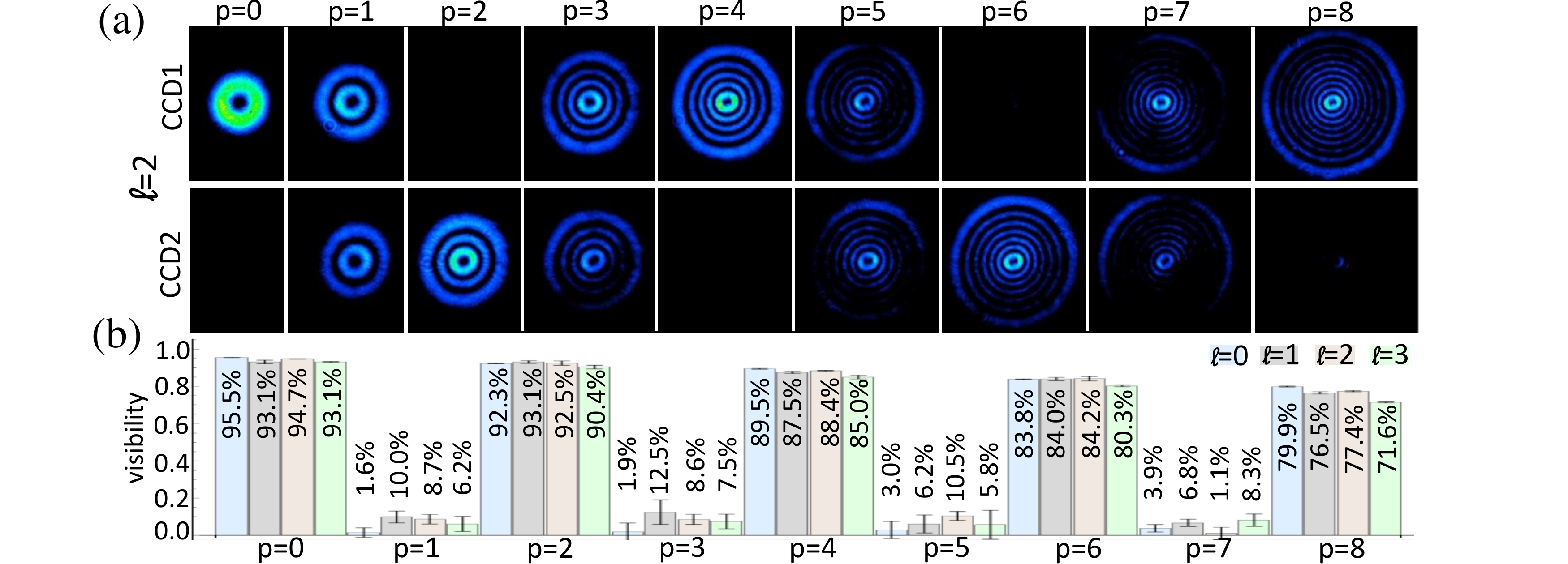}
\caption{Experimental results for $p$ and $\ell$ modes in a $\Delta\varphi_{g}=\frac{\pi}{4}$ interferometer. \textbf{A}: Here we show an example for $\ell$=2, $p$=0-8. From the two CCD outputs, we clearly see modes with even $p$ constructively/destructively interfere, while modes with odd $p$ randomly split. \textbf{B}: The corresponding visibilities are consistently larger than 75\%. Even modes with up to $p$=8, $\ell=2$ (which has a mode order of $m=(2p+\ell+1)=19$) can be interfered well beyond 75\%. The errors stand for statistical uncertainties calculated from 10 independent measurement data points.}
\label{fig:QuarterPiSort}
\end{figure*}

To ensure interference even with slightly different optical pathlength introduced by the lenses in one arm, we use a $\lambda$=810nm laser with a sufficiently long coherence length. To create higher-order Laguerre-Gauss modes with a high-quality, we use a phase-only Spatial Light Modulator (SLM), and apply the method to generate such modes suggested in \cite{bolduc2013exact}.

Radial modes are dependent on the beam waist, thus it is intuitively clear that the interferometer requires that the input beam has specific properties to give correct Gouy phases (in our case, $\lambda$=810nm, $w_0$=1mm, $z$=0m). For long-term stability, we build the interferometer in a double-path Sagnac configuration (see Supplementary for details on the experimental setup). After the interferometer, we split the output beams and direct them to two CCDs (which we use to image the output modes) and two power-meters (which we use to measure the interference visibilities).

\textbf{Experimental results for $\frac{\pi}{2}$} -- In order to produce $\Delta\varphi_{g}=\frac{\pi}{2}$, we use L1=500mm, L2=40mm, L3=300mm, D1=560mm, D2=343mm and a corresponding free-space propagation length in path B of D=D1+D2=903mm. In Fig. \ref{fig:halfPisort}A-B, we show the results for sorting the radial $p$ modes from $p$=0 to $p$=3 (with $\ell$=0, 1, 2 and 3). The visibilities are all above 75\%, and 83.5\% on average. In Fig. \ref{fig:halfPisort}C-D, we show the sorting of $\ell$ modes, every even mode is sorted, while every odd mode is equally separated in both output ports. We measure $\ell=0-6$ for $p=0$ ($\ell=0-6$ with $p$=1, 2 and 3 can be seen in the Supplementary). The sorting efficiency is very high, leading to visibilities beyond 85\%. Intuitively, one would expect a lower visibility for higher order modes. Instead, we observe here the highest visibility for the $p=1$ radial mode. This has two reasons: Deviations of the experimentally generated input beam $q_{in}$ (beam waist $w_0$ and focus position $z$) from the ideal beam; and small errors in the distances between the three lenses. These misalignments lead to a slightly different accumulated Gouy phase difference, with the effect that higher order modes are sorted less efficient. Additionally, they also lead to different complex $q$ parameters for the two paths A and B, and therefore to Newton rings which lower the interference visibility. None of the above two reasons are of fundamental nature and can be overcome with carefully designing and manufacturing the interferometric device.

In order to show that our proposed method is also capable for possible quantum applications, we investigate the device's capability of sorting coherent superpositions of different radial and OAM modes. In Fig.~\ref{fig:Superposition}, we show the measurement results for coherent superpositions of LG modes with different orthogonal modes. In particular, we show the coherent superposition of $\psi=\frac{1}{\sqrt{2}}\left(\text{LG}_{2,0}+\text{LG}_{0,2}\right)$ and $\psi=\frac{1}{\sqrt{2}}\left(\text{LG}_{1,1}+\text{LG}_{2,1}\right)$. High quality separation of the superposition is clearly visible.

\textbf{Experimental results for $\frac{\pi}{4}$} -- The device explained above can sort modes with even and odd $p$ values. However, it can not separate two even $p$ values. With simple adjustments of the lenses (D1=506mm and D2=326mm), we are able to perform a $\Delta\varphi_{g}=\frac{\pi}{4}$ phase shift which can separate even and odd $p$/2 modes, such as $p$=0 and $p$=2 -- as shown in Fig. \ref{fig:QuarterPiSort}A. In Fig. \ref{fig:QuarterPiSort}B, the sorting visibilities of LG modes with $\ell=2$ and $p$=0 to $p$=8 are shown. Every odd mode is equally separated, while the even modes are sorted. All visibilities are beyond 85\% for $p\leq4$ and larger than 75\% for all modes smaller or equal to LG$_{p=8,\ell=2}$.

\textbf{Conclusion} -- We have identified and realized a method to interferometrically sort higher-order spatial Gaussian modes using accumulated Gouy phases. Particularly, it allowed us to experimentally sort radial Laguerre-Gaussian modes. It is also possible to cascade several $p$-mode sorters to increase the independently accessible modes, or to combine it with OAM-mode sorters \cite{leach2002measuring} to access the complete set of spatial modes. This can readily be used for multiplexing and demultiplexing technologies in high-speed classical communication schemes. Our presented method can also be used in quantum optics as an in principle lossless two-input two-output device. This is in direct analogy to the polarizing beam splitter which is the workhorse for multi-photon qubit entanglement experiments \cite{pan2012multiphoton}. The $p$-mode sorter expands the toolbox to manipulate higher-dimensional quantum states \cite{zhang2016engineering} and generate multi-photon high-dimensional entanglement \cite{malik2016multi, erhard2017experimental}. It can also be used to create controlled quantum gates (e.g. CNOT gates) exploiting hybrid systems involving radial spatial modes. Having access to the radial modes would allow quantum teleporation of multiple degrees-of-freedom of a photon \cite{wang2015quantum}. In particular it could enable to teleport the complete quantum information encoded in the 2-dimensional transverse plane of a single photon. Two basic building blocks for arbitrary high-dimensional unitary transformations are the generalized X- and Z-gate \cite{babazadeh2017high,asadian2016heisenberg}. Our demonstrated $p$-mode sorter implicitly uses a generalized Z-gate (the lens system). The high-dimensional X-gate requires individual access to different parities (which we have demonstrated here) as well as a mode shifter. Thus the last missing experimental tool to generate arbitrary unitary transformations in the radial degree of freedom is the possibility to shift modes by a constant value. The implementation of such a mode shifter remains an important open question.

\textit{Note:} While finishing this manuscript, we learned about a similar research project \cite{zhou2017sorting}. There, the authors solve the question for p-mode dependent phase in a different way: They use a \textit{fractional Fourier transform} for which the p-modes are eigenfunctions.

\section*{Acknowledgements}
X.G. thanks Lijun Chen for support. This work was supported by the Austrian Academy of Sciences (\"OAW), by the Austrian Science Fund (FWF) with SFB F40 (FOQUS). XG acknowledges support from the National Natural Science Foundation of China (No.61272418) and its Major Program (No. 11690030, 11690032).

\bibliographystyle{unsrt}
\bibliography{refs}

\clearpage

\widetext
\begin{center}
\textbf{\Large Supplemental Materials}
\end{center}

\section{Appendix I. Detailed experimental scheme}
\begin{figure}[htbp]
\centering
\includegraphics[width=\textwidth]{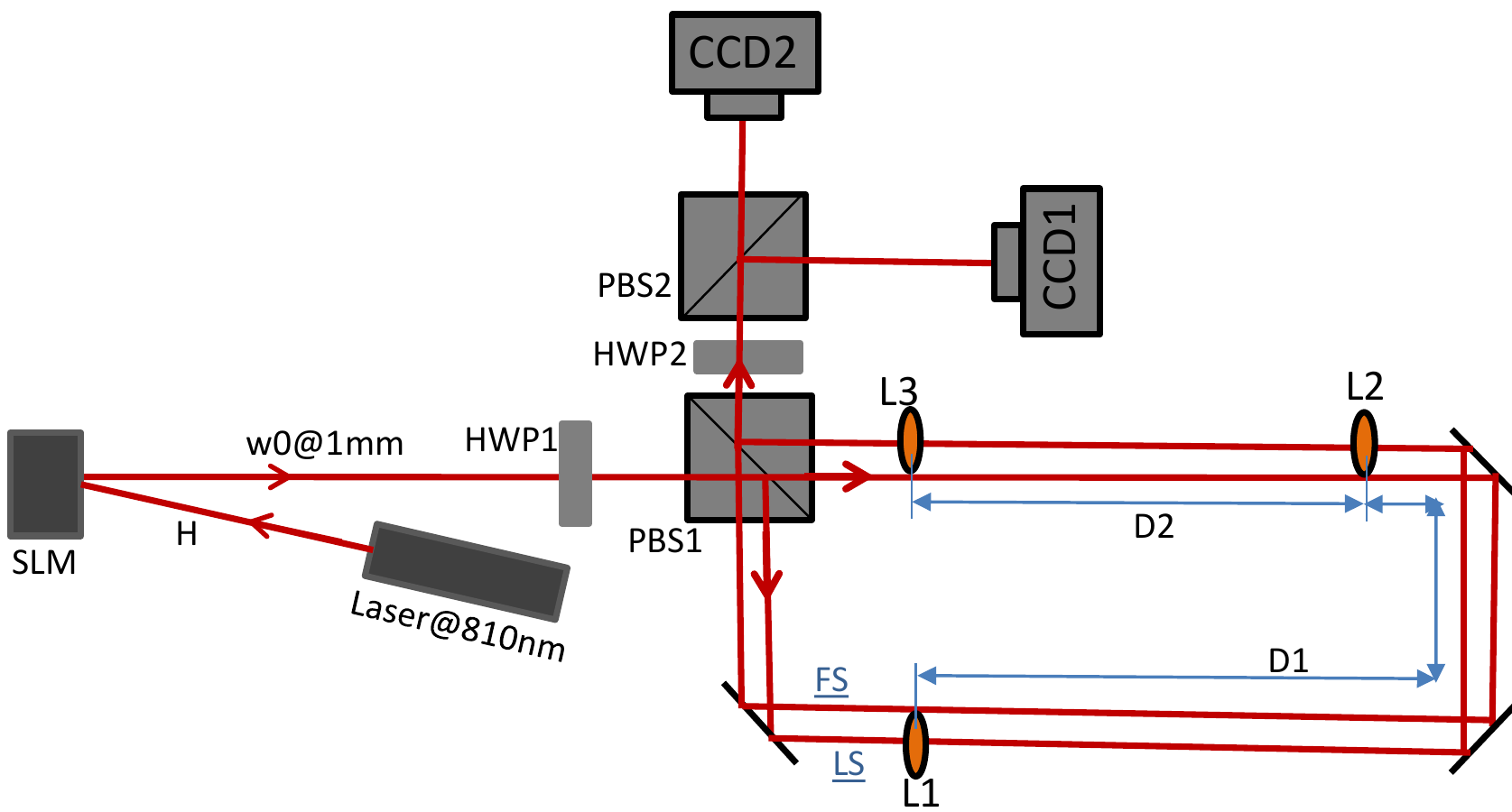}
\caption{\textbf{Experimental setup}: We use a horizontally polarised, long coherence length laser with a wavelength of $\lambda=810$nm. We use the Spatial Light Modulation (SLM) to create a gaussian beam waist of $w_{0}$=1mm and to generate higher-order spatial modes. We utilize a half-wave plate (HWP1) to change the polarization in order to modulate the intensities for the two arms in the subsequent interferometer. The beam is splitted into two arms by a polarizing beam-splitter (PBS1). The goal is to sort higher-order spatial modes in an interferometric way, by exploiting the difference of the accumulated Gouy phase in the two arms. To keep the interferometer stable for long time, we build the interferometer in a Sagnac configuration. In the interferometer, we insert three lenses (L1, L2, L3) in one of the arms (LS) while the other arm consists only of free space propagation (FS). For $\Delta\varphi_{g}=\frac{\pi}{2}$, the focus lengths are L1=500mm, L2=40mm, L3=300mm and distances between these lenses are D1=560mm, D2=343mm respectively.  In order to observe the interference between the two arms, we use HWP2 (H $\to$ D, V $\to$ A) and PBS2. In the end, we detect the interference with CCDs (CCD1, CCD2) and power meters (not shown).}
\label{fig:setupSI}
\end{figure}

The detailed experimental configuration is described in Fig. \ref{fig:setupSI}.

\newpage

\section{Appendix II. visibility analysis for experimental results }
\begin{figure}[htbp]
	\centering
	\includegraphics[width=0.45\textwidth]{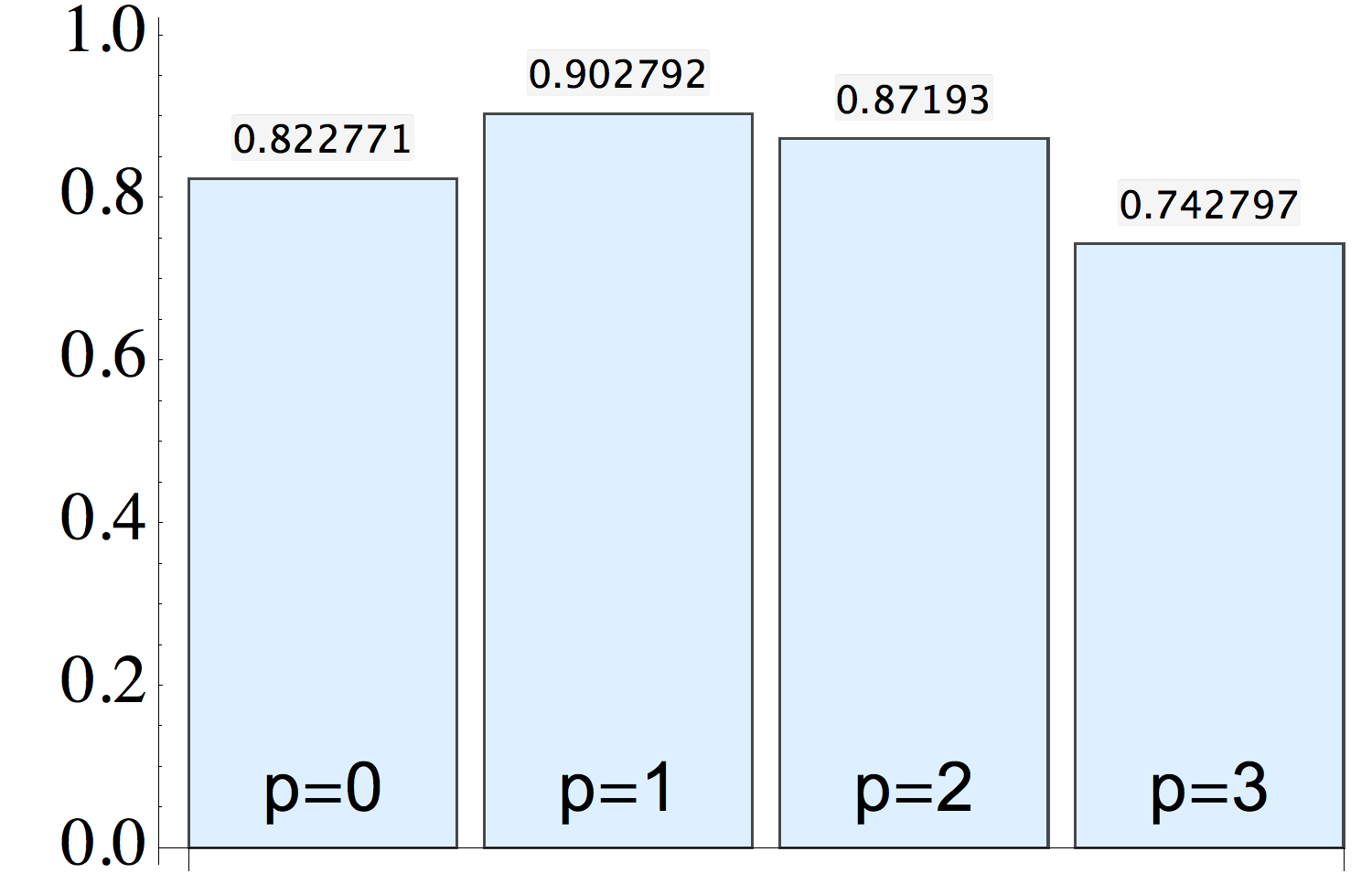}
	\caption{Theoretical prediction of the visibilities for different $p$-modes and $\ell=0$.}
	\label{fig:theory-p-mode-vis}
\end{figure}

Here we investigate the experimental results presented in Fig. 2 and 4 (main text) in detail and explain the observed results. In Fig. 2 we show the sorting visibilities for $\text{LG}^\ell_p$ with different $\ell=0,1,2,3$ and $p=0,1,2,3$. Theoretically, one would expect perfect visibilities of 1 and if the visibilities deviate from 1, then the intuitive guess would be that higher order modes generally show a lower visibility. In contrast, our results show the highest visibility for all $\ell=0,1,2,3$ modes with $p=1$. We can explain this specific structure by theoretically investigating our experiment in detail. Therefore we model the $\text{LG}^\ell_p$ mode in terms of the $q$ parameter
\begin{align*}
	\text{LG}^\ell_p=\frac{1}{w_z(q)}\exp \left(-\frac{r^2}{w_z(q)^2}\right) \left(\frac{\sqrt{2} r}{w_z(q)}\right)^{\left| \ell\right| } L_p^{\left| \ell\right| }\left(\frac{2 r^2}{w_z(q)^2}\right)    \exp \left(-i \left(-\psi_z(q,\ell,p)+\frac{k r^2}{2 R_z(q)}+k Re(q)+\ell \phi \right)\right),
\end{align*}
with the Rayleigh range $z_R(q)=Im(q)$, the beam waist $w_z(q)=\sqrt{-\frac{\lambda }{\pi  Im\left(\frac{1}{q}\right)}}$, radius of curvature $R_z(q)=\frac{1}{Re\left(\frac{1}{q}\right)}$, the wave vector $k=\frac{2 \pi }{\lambda }$ and $\lambda$ denoting the central wavelength of the beam and the Gouy phase $\text{$\psi_z$}(q,\ell,p)=(\left| \ell\right| +2 p+1) \tan ^{-1}\left(\frac{Re(q)}{Im(q)}\right)$. We simplify our system to the experimental setup shown in Fig. 1. The incoming beam $q_\text{in}$ is splitted up at a 50/50 beam-splitter. The freespace and lens transformations are then calculated according to the ABCD matrix formalism. The second 50/50 beam-splitter recombines the two beams. We then calculate the visibility according to
\begin{align*}
	V=\frac{I_a-I_b}{I_a+I_b},
\end{align*}
where $I_{a,b}$ denotes the integrated intensity distribution in the respective path $a,b$ after the interferometer.
The interference visibility for different $p$-modes is determined by the Gouy-Phase difference for the two paths in the interferometer $\Delta\varphi_{g}$, which crucially depends on the $q$-parameter. The $q$ parameter is changed in one path with three subsequently placed lenses (L1,L2,L3) with distances D1 and D2 between them. Depending on the precise focal length values of the three lenses and the distances between them, one can adjust for a given input beam $q_\text{in}$ (with a beam waist of $w_0=1mm$ at the first lens L1) a Gouy-phase difference of $\pi/2$. In our case this is for L1=500mm, L2=40mm, L3=300mm, D1=560mm and D2=343mm. For these values one expects perfect visibilities. Setting exactly these values is difficult in the experiment. We measured that the experimentally created Gaussian input beam waist is $w_{0,exp}=(0.96\pm0.004)mm$ and the $z=0$ position is about $z_{0,exp}=(-200\pm100)mm$ shifted. Slight deviations of the two lens distances $\text{D1}+3mm$ and $\text{D2}+3mm$ leads to the visibilities shown in Fig. \ref{fig:theory-p-mode-vis}.
Note, that here we also adjusted an additional mode independent phase $\Delta\phi$ between the two paths. We adjusted this phase $\Delta\phi$ in the experiment with a piezo controlled mirror for the $p=2$ mode. We simulated exactly this also in the theoretical investigation. In conclusion, two effects are responsible for lowering the interference visibilities in our experiment. First, slightly different input Gaussian beam parameters $q$ pick up a different Gouy-phase shift. This effect can actually be compensated with the mode-independen phase shift $\Delta\phi$, although only for one specific mode. The second effect is that a different input parameter $q_\text{in}$ into the lens transformation leads to a different output parameter $q_\text{out}$. This leads for the Gaussian beam to the well known Newton rings and additionally lowers the visibility. Both effects are not a fundamental limitation and can be overcome by carefully designing and manufacturing the interferometer.

\newpage

\section{Appendix III. Experimental results for sorting $\ell$ modes with $\Delta\varphi_{g}=\pi/2$}

In the main main text (Fig. 2C) we demonstrate the sorting of $\ell$ mode from 0 to 6 with $p$=0, using an accumulated Gouy-phase change of $\Delta\varphi_{g}$=$\pi/2$. Here we change the radial mode number $p$ from 1, 2, 3 to sort $\ell$ from 0 to 6 in the same way.

\begin{figure}[htbp]
\centering
\includegraphics[width=\textwidth]{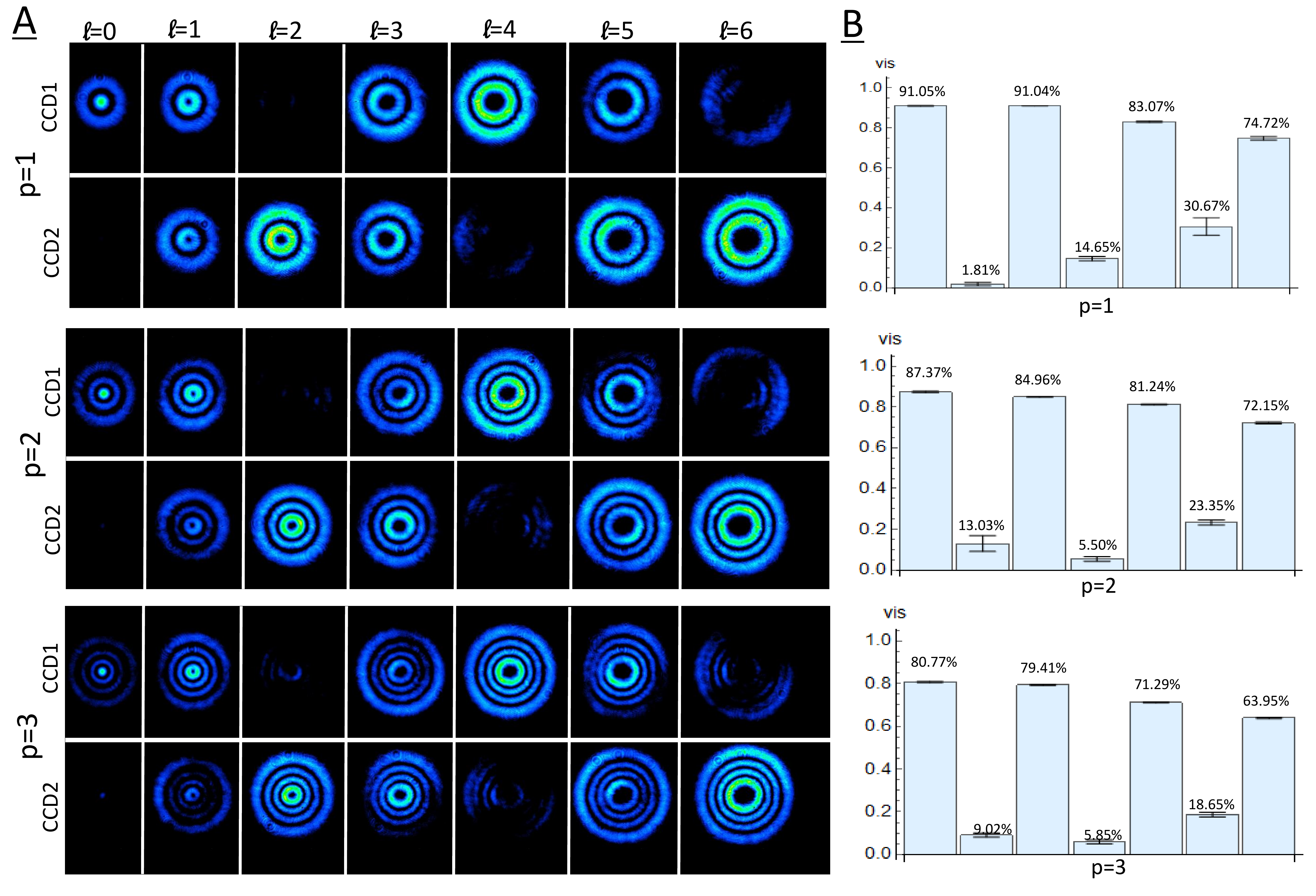}
\caption{Experimental results of sorting $\ell$ modes with constant $p$ mode using $\Delta\varphi_{g}=\frac{\pi}{2}$. \textbf{A}: Top left is for $p$=1,$\ell$=0-6. From the two CCD outputs, we clearly see that $\ell$=0,2,4... modes be separated into the states with $\ell$=0,4...(i.e., ($\ell$ mod 4)=0) and $\ell$=2,6...(i.e., ($\ell$ mod 4)=2). Middle left is for $p$=2, $\ell$=0-6; bottom left is for $p$=3, $\ell$=0-6. \textbf{B}: The corresponded measured visibility for $p$=1, $\ell$=0-6; $p$=2, $\ell$=0-6; $p$=3, $\ell$=0-6. Error bars are statistical error by analyzing ten random data points.}
\label{fig:Psort}
\end{figure}

\newpage

\section{Appendix IV. Experimental results for sorting $p$ modes with $\Delta\varphi_{g}=\pi/4$}

In the main main text (Fig. 4) we demonstrate the sorting of $p$ modes with $\ell=2$, using an accumulated Gouy-phase change of $\Delta\varphi_{g}$=$\pi/4$.  Here we show additional experimental results for $\ell$=0,1,3 and $p$ changes from 0 to 8.

\begin{figure}[htbp]
\centering
\includegraphics[width=\textwidth]{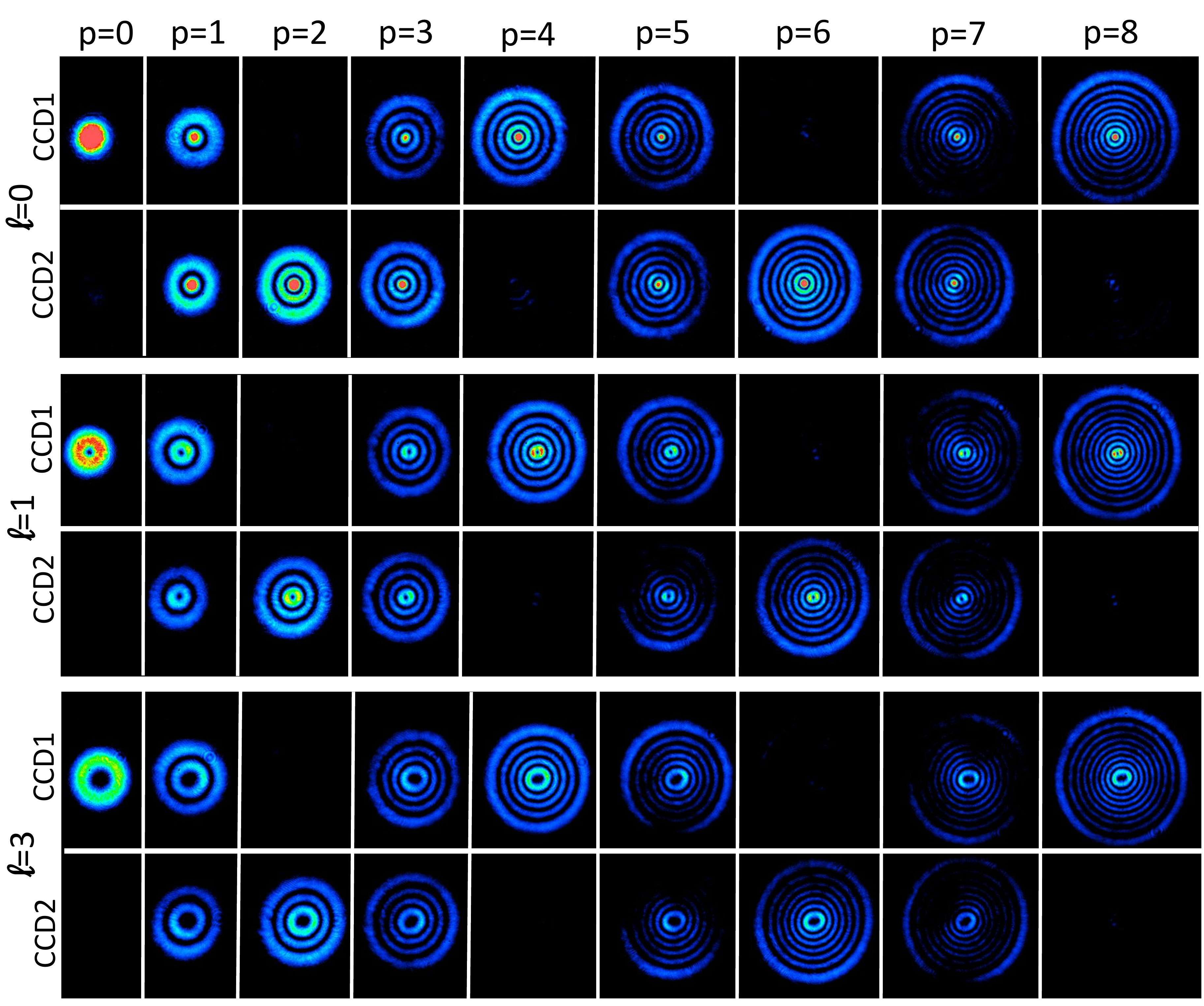}
\caption{Experimental result for sorting $p$ modes with constant $\ell$ value by $\Delta\varphi_{g}=\frac{\pi}{4}$. From the two CCD outputs, we clearly see $p$=0,4,8 and $p$=2,6 be separated.}
\label{fig:Lsort}
\end{figure}

\newpage
\section{Appendix V. List of Lenses used in the automated Search for the experimental Configurations}

We used the following set of lenses, conviniently available for instance at Thorlabs, in our automated search for experimental configurations:

\begin{table}[h]
  \centering
  \begin{tabular}{|c|c|c|c|c|c|c|c|c|c|c|c|}
    \hline
    \multirow{2}{*}{ focal length (mm) } & 25.4&30.0&35.0&40.0&50.0&60.0& 75.0&100.0&125.0&150.0&175.0 \\ \cline{2-12}
       & 200.0 & 250.0 &300.0 &400.0 &500.0 &750.0 &1000.0 &-50.0 &-75.0 &-100.0 & \\ \hline
     \hline
    \end{tabular}
   \caption{focal length of Bi-Convex and Bi-Concave lenses from Thorlabs}
   \label{tab:lenseslist}
\end{table}

\section{Appendix VI. Experimental configurations for various $\Delta\varphi_{g}=\pi/n$ phase differences}

From the computational search for experimental configurations, we found simple configurations using standard lenses to perform phase differences of $\Delta\varphi_{g}=\pi/n$ (with $2\leq n\leq 8$). All configurations consist of only three lenses, many other types of phase can be found in this way.

\begin{table}[h]
    \centering
    \begin{tabular}{|c|c|c|c|c|c|}
    \hline

    \multirow{2}{*}{$ \frac{\pi}{n} $} & \multirow{2}{*}{$\Delta\varphi_{g}$} & \multicolumn{2}{|c|}{\textbf{Configuration}} & \multicolumn{2}{|c|}{  Visibility  }\\ \cline{3-4}
    &  & \multicolumn{1}{|c|}{ L1, L2, L3 (mm) } & \multicolumn{1}{|c|}{ D1, D2 (mm) } & \multicolumn{2}{|c|}{ $p$ : $vis \%$}  \\ \hline

    2 & -0.501$\pi$  & 500    40    300 & 555.6    339.0 & 0: 99.96  & 2: 99.93\\
    3 &  0.333$\pi$  & 400    30    300 & 414.3    316.2 & 0: 99.99  & 3: 99.98\\
    4 &  0.251$\pi$  & 500    40    300 & 502.7    320.9 & 0: 99.91  & 4: 99.29\\
    5 &  0.199$\pi$  & 300    35    400 & 313.5    395.8 & 0: 99.98  & 5: 98.55\\
    6 & -0.166$\pi$  & 300    75    200 & 485.3    316.2 & 0: 99.83  & 6: 97.89\\
    7 & -0.143$\pi$  & 150    75    300 & 266.0    155.4 & 0: 99.97  & 7: 96.57\\
    8 &  0.124$\pi$  & 300    30    250 & 294.6    253.6 & 0: 99.80  & 8: 98.12\\
     \hline
    \hline
    \end{tabular}
    \caption{The configuration of lens systems for experimental setup with different Gouy phase $\Delta\varphi_g$. Each of the configuration uses three lenes in one arm. The expected interference visibility for $p$=0 and higher-order $p$ modes is shown in the right column. Note that in the experiment described in the main text, we used plano-convex lenses, thus the distances D1 and D2 are slightly different than in this table. For these configurations, the beam focus lies at the lens L1 with a beam waist of 1 mm and a wavelength of 810 nm.}
   \label{tab:configuration}

\end{table}

\end{document}